# Application of digital regulated Power Supplies for Magnet Control at the Swiss Light Source

A. Lüdeke, PSI, Switzerland

*Abstract*

The Swiss Light Source (SLS) has in the order of 500 magnet power supplies (PS) installed, ranging from from $3\,A/20\,V$ four-quadrant PS to a $950\,A/1000\,V$ two-quadrant $3\,Hz$ PS. All magnet PS have a local digital controller for a digital regulation loop and a $5\,MHz$ optical point-to-point link to the VME[1] level. The PS controller is running a pulse width/pulse repetition regulation scheme, optional with multiple slave regulation loops. Many internal regulation parameters and controller diagnostics are readable by the control system. Industry Pack modules with standard VME carrier cards are used as VME hardware interface with the high control density of eight links per VME card. The low level EPICS[2] interface is identical for all 500 magnet PS, including insertion devices. The digital PS have proven to be very stable and reliable during commissioning of the light source. All specifications were met for all PS. The advanced diagnostic for the magnet PS turned out to be very useful not only for the diagnostic of the PS but also to identify problems on the magnets.

## 1 INTRODUCTION

The SLS design required a large number of magnet PS with a wide power range (see figure 1) and a variety of different features. For the booster a $3\,Hz$ sine wave ramping mode is needed. The beam-based alignment option requires the ability to introduce an individual $5\,Hz$ wobbling of the current for each of the 192 quadrupole magnets of the Linac, transfer-lines and storage ring. The ring orbit feedback requires $1\,kHz$ control bandwidth with more than a 17 bit ($15\,ppm$) granularity of the corrector current set-points.

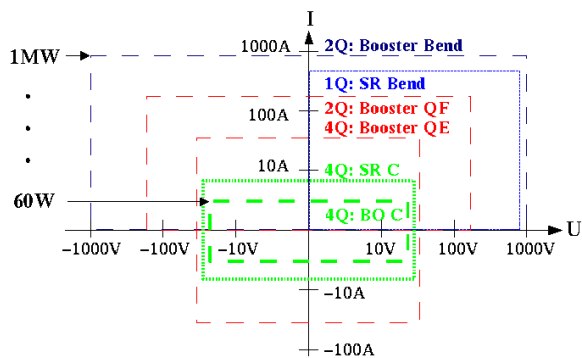

Figure 1: Operation quadrants range of the SLS magnet PS

---

[1] Versa Modular European
[2] Experimental Physics and Industrial Control System

The PSI division Electrical Engineering, responsible for the delivery of the PS to the SLS project chose a fully digital controlled, uniform solution for all magnet PS at the SLS. The prototypes of the PS were developed at PSI but the series were manufactured by outside companies.

## 2 CONTROL OVERVIEW

The overall control hardware scheme is shown in figure 2. Each PS has a digital regulation loop build by a digital sig-nal processor (DSP) and Pulse width modulation (PWM) card, a power converter, a current measurement (DCCT) and a precise analog to digital- and digital to analog converter (ADC/DAC) card. A point-to-point optical fiber link connects the PS controller with an industry pack module hosted on a VME carrier board. An industry pack (IP) module, developed by the Accelerator Control Group at PSI, serves two PS, each one with one bidirectional $5\,MHz$ link. Four IP modules fits on each IP carrier VME card. Therefore even with the point to point high bandwidth link a high control density of eight PS per VME card can be reached. On average 16, at maximum 24 PS are controlled by one crate.[3] The VME side hardware costs are approx. 180 $ per PS link. The PS controller card and the ADC/DAC card were manufactured by outside companies for about 2000 $ per PS, mainly for the controller. Plastic Optic Fiber were used for the link, a very cheap equivalent to optical fiber for short (up to $50\,m$) connections.

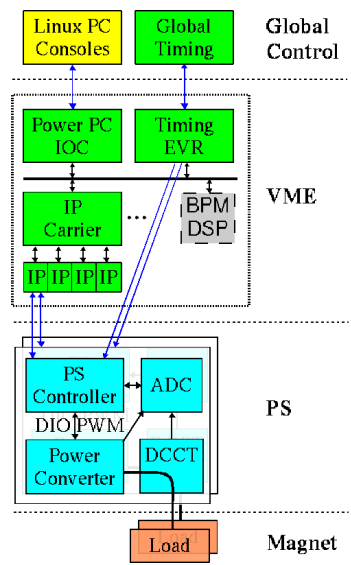

Figure 2: The control hardware scheme

## 3 POWER SUPPLY CONTROLLER

The choice of a digital regulation loop reduces the drift problems of a PS to mainly one source: the analog digital conversion. A solution with two true 16 bit ADCs of $100\,kHz$ bandwidth each and self calibration capability is used to generate a 17 bit plus sign analog signal for the $50\,kHz$ control loop. Additional ADCs on the ADC/DAC card are used to measure voltages of the power converter. Two programmable DAC outputs on the card can be used for local diagnostics. Each analog register value of the controller can be assigned to each DAC with a programmable offset and scaling. This for example makes it possible to look at fluctuation of the 14th-20th bit of the read-current at a constant set current.

The high desired precision of the PS were reached by applying new methods on the PWM loop: a rounding correction and a pulse repetition modulation (PRM). Some PS have several digital controllers connected in a master and slave chain via a 30 $MBaud$ backplane link of the 60 $MHz$ DSP cards. For example the 1 $MVA$ booster PS has 7 regulation loops on 4 DSP cards.

The optical link is controlled by a FPGA on both ends: on the DSP card and on the IP module. The protocol supports read/write access from the IP modules (bus-master) to 256 registers of 32 bits of the controller. A throughput rate of 10k float values per second is guaranteed and a maximum latency of 30 $\mu s$ to set a current. Current waveforms and controller programs can be downloaded to either the DSPs SRAM or the 8 $MBit$ Flash EEPROM.

The industry pack module has a special high priority register to allow a direct VME access to the PS without going through the PSC driver. This is needed for the fast orbit feedback, where a local VME DSP board writes set currents to the corrector PS through the VME bus with a rate of 1 $kHz$.

Each Controller has an additional optical trigger input. The trigger input is used by the controller to start a programmable current waveform. A precise $50\,kHz$ clock on the DSP board clocks the set-point steps, one set-point each $80\,\mu s$ and three interpolation steps. An offset and a scaling of the waveform can be controlled independently of the actual waveform.

A more detailed description of the digital PS controller can be found in [1].

## 4 SOFTWARE LAYER

The software functionality is organized in several layers as shown in figure 3.

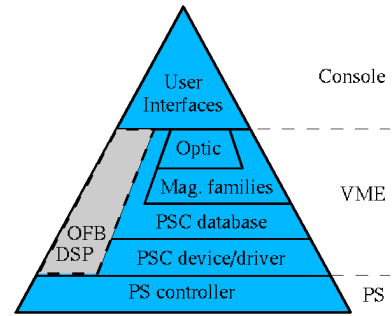

Figure 3: The control software scheme

### 4.1 PS Controller Software

The regulation loop and the control of fast current waveforms is handled by the PS controller. All the internal regulation parameters can be read from there and many self tests are performed here. The software of all PS controller DSPs is in general identical and differs only by regulation parameters in the Flash EEPROM.

### 4.2 VME Software

The next two layers – the PS control (PSC) EPICS driver / device support and the PSC EPICS database – are also identical for all PS. The EPICS templates have a total of 120 records. A third of these are connected to registers of the PS controller. Beside of the standard parameters like the switch, set and read-back current and the status of the PS there are parameters for an enhanced fault diagnostic like the average load resistance calculated by the controller. Another 40 channels are used for diagnostics of the optical fiber link and were mainly needed for the development. Only three channels are used frequently here: the flags for a broken transmit link[4], a broken receive link and the flag for local control of the controller. For the download of DSP software and current waveforms 25 channels are used, mainly just to hold the data to download. The other channels are soft channels for extra features, like to control the initialization after booting the VME, a proper hysteresis handling, advanced alarm-handling and for VME controlled slow current waveforms. The latter ones are needed for a synchronized ($< 1ms$ Jitter) slow (seconds to minutes) ramping of an arbitrary set of magnet PS, for example local orbit bumps or an energy ramping of the storage ring.

The storage ring has individual PS for each of the 174 quadrupoles. From the optics they are grouped into 31 families with an identical theoretical current. These families are controlled by another EPICS database. All public functionality of an magnet PS is reproduced by this magnet family database. Therefore the same user interfaces can be used for a magnet family than the one used for individual magnet PS control. Each family member PS can have an offset or a scaling relative to the set value for the family. This feature is used to compensate for measured local distortions of the focusing.

---

[3] Up to 160 PS can be controlled by one VME crate, if a reduced control bandwidth is acceptable.

[4] the PS controller detects this and sends a message to the IP module

Since the storage ring still has 31 quadrupole families and 9 sextupole families a further reduction of the adjustable parameters was desired. Therefore the EPICS database "Optic" reduces the number of control parameter to basically five physical quantities: the Energy, a horizontal and a vertical tune shift and a horizontal and a vertical shift of the chromaticity. A theoretical set of magnet currents and the matrices to compute the actual set currents from the physical quantities are off-line generated from a beam dynamics modeling program and downloaded to the optic database. Saturation effects are calculated by the model and are taken into account in the database by individual gradients of the families.

The software of the orbit feedback directly accesses the PS controller level without disturbing the PS device / driver.

### 4.3 Console Software

The standard user interface for magnet control is implemented as a special widget in the generic tcl/tk[5] program "panel.tcl" to control a list EPICS devices. The main pa-rameters are: a set current, read current, compare flag, hysteresis state and status button. The compare flag shows an alarm if set- and read current differ more than the limits given by the specifications. The hysteresis state shows if the magnet is still on its nominal hysteresis branch or if it would need to be standardized by the individual current cycle. Further status information and controls can be accessed by the more extensive pscStatus application launched by the "Status" button. This graphical interface is customized to the individual functions of the specific power supply.

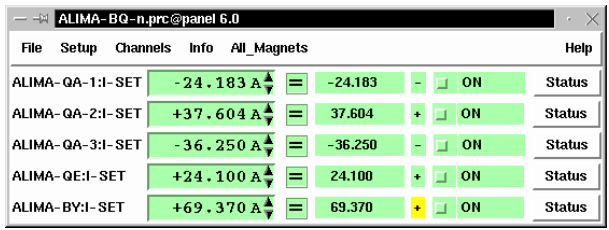

Figure 4: The generic magnet user interface

Although the panel.tcl program has configuration files for all magnet PS and all magnet families, it is mainly used to control magnets in the Linac and the transfer lines and for the DC offset of the booster ramps. As mentioned above, the storage ring quadrupole and sextupole magnets are controlled by optic parameters only. The booster magnet current ramps are generated and downloaded by a ramp editor and Corrector magnets are rather controlled by the orbit correction applications. The generic interface is mainly used for fault handling of these PS.

---

[5]The Tool Command Language (tcl) and its graphical ToolKit (tk)

## 5 EXPERIENCES DURING THE SLS COMMISSIONING

The first 11 prototypes were ready for Linac Commissioning in November 1999. All layers of hard- and software had just a limited set of functionality then (no waveform handling, one link per IP module, ..) but the setup worked fine and all nessesary features were availiable. The Linac magnet powersupplies were used from then on as a testbed for continous further developments on the hard- and software, without disturbing the comissioning progress.

Another 125 PS were operational for the Booster from July 2000 on. At that time also several parts of the linac hardware were replaced by new releases of PS controllers and IP modules. Soft- and hardware worked reliable during booster comissioning. Only about 1% of downtime was encountered due to failures of power supplies, PS controllers or PS VME systems. The main source here were defective DC-DC converter on the PS controller board. The manufacturer later acknowledged a production error and the hardware was replaced.

Another 348 PS were operational for the storage ring commissioning starting December 2000. A new IP FPGA software was used with enhanced diagnostic capability and some enhancements were done to the DSP software. On VME level an alarm on the magnet resistance was introduced when short circuits were detected on some sextupole windings. This helped also to find a cabeling error which lead to two interchanged quadrupoles in the storage ring.

The electromagnets of the insertion devices will also be powered by digital PS. The first electromagnetic device UE232, consisting of two 4.4 meter electro-magnetic crossed field undulators, started comissioning in August 2001. Even this device has the same PSC EPICS database and device / driver, additional functions are added by superordinated EPICS databases.

## 6 CONCLUSION

The choise of the new technology of digital regulated power supplies had no drawbacks for the commissioning of the swiss light source. No mentionable extra time was needed for the commissioning of the power supplies and even the prototypes allowed reliable operation. The total development time was rather short since one solution could be used for all of the 500 power supplies. This is valid for hardware and software development of all layers.